\documentclass[aps,prb,twocolumn,superscriptaddress,showpacs]{revtex4-1}

\usepackage{ulem}
\usepackage[utf8]{inputenc}
\usepackage{graphicx}
\usepackage{dcolumn}
\usepackage{bm}
\usepackage{hyperref}
\usepackage{amsmath}
\usepackage{mathtools}
\usepackage[dvipsnames]{xcolor}

\newcommand{\qperp}{q}
\newcommand{\bqperp}{{\bf{q}}}

\begin{document}

\preprint{APS/123-QED}

\title{Finite-size effects in hyperuniform vortex matter}

\author{Roc\'{i}o Milagros Besana$^{1,2,3}$, Federico El\'{i}as$^{1,3,4}$, Joaqu\'{i}n Puig$^{1,2,3}$, Jazm\'{i}n Arag\'{o}n S\'{a}nchez$^{1,2,3}$, Gladys Nieva$^{1,2,3}$, Alejandro Benedykt Kolton$^{1,3,4}$, and Yanina Fasano$^{1,2,3,5}$ }%
\address{$^1$Low Temperatures Lab, Centro At\'{o}mico Bariloche, CNEA, Argentina.}
\address{$^2$Instituto de Nanociencia y Nanotecnología, CONICET-CNEA, Nodo Bariloche, Argentina.}
\address{$^3$ Instituto Balseiro,
CNEA and Universidad Nacional de Cuyo,
Bariloche, Argentina}
\address{$^4$ Condensed Matter Theory group, Centro At\'{o}mico Bariloche, CNEA, Argentina.}
\address{$^5$ Leibniz Institute for Solid State and Materials Research,  Dresden, Germany}

\date{\today}

\begin{abstract}
Novel hyperuniform materials are emerging as an active field of applied and basic research since they can be designed to have exceptional physical properties. This ubiquitous state of matter presents a hidden order that is characterized by the density of constituents of the system being uniform
at large scales, as in a perfect crystal, although they can be isotropic and disordered like a liquid. In the quest for synthesizing hyperuniform materials in experimental conditions, the impact of finite-size effects remains as an open question to be addressed.  We use vortex matter in type-II superconductors as a toy model system to study this issue. We previously reported that vortex matter nucleated in samples with point disorder is effectively hyperuniform and thus presents  the interesting physical properties inherent to hyperuniform systems. In this work we present experimental evidence that on decreasing the thickness of the vortex system its hyperuniform order is depleted.
By means of hydrodynamic arguments we show that
the experimentally observed depletion can be associated to two crossovers that we describe within a hydrodynamic approximation. The first crossover length is thickness-dependent and separates a class-II hyperuniform regime at intermediate lengthscales from a regime that can become asymptotically non-hyperuniform for large wavelengths in very thin samples. The second crossover takes place at smaller lengthscales and marks the onset of a faster increase of density fluctuations due to the dispersivity of the elastic constants. Our work points to a novel mechanism of emerging hyperuniformity controlled by the thickness of the host sample, an issue that has to be taken into account when growing hyperuniform structures for technological applications.

\end{abstract}

 \maketitle

\section*{Introduction}

Hyperuniformity is a ubiquitous asymptotic structural property shared by many physical, biological and mathematical systems. In hyperuniform systems the distribution of their constituents may be ordered or disordered at first sight but  presents a hidden order characterized by
a suppression of density fluctuations at large scales.~\cite{Torquato2003}
More specifically, density fluctuations decrease with increasing wavelengths and strictly vanish in the large wavelength limit. In many physical systems, the trade-off between the interaction among constituents favouring the formation of a lattice, and the interplay with disorder, may result in the nucleation of disordered hyperuniform structures. Disordered hyperuniform materials are endowed with a novel phenomenology that goes against conventional wisdom on the effect of disorder in systems of interacting objects.~\cite{Man2013,Torquato2018,Chen2018} For instance, disordered hyperuniform two-dimensional silica structures present a closure of the electrical conduction gaps, producing a lowering of resistance in comparison with ordered structures.~\cite{Zheng2020} Another example are disordered hyperuniform materials that possess complete photon conduction gaps for short wavelengths, blocking all directions and polarizations of light at high frequencies.~\cite{Man2013,Florescu2009,Froufe2016} Thus, novel hyperuniform materials are emerging as an active field of applied and basic research since they can be designed to have exceptional physical properties.

Hyperuniform patterns typically
present a structure factor $S(\mathbf{q})= S(q_{\rm x},q_{\rm y})$ that algebraically tends to zero for reciprocal space wave-vectors $\mathbf{q} \rightarrow 0$.~\cite{Torquato2003}
The structure factor is the squared-modulus of the Fourier transform of the local
density fluctuations, namely $S(\mathbf{q})=  |\hat{\rho}(q_{\rm x},q_{\rm y},z=0)| ^{2}$.
Since at thermal equilibrium the value of
$S(\mathbf{q})$ for $q=0$ is proportional to the compressibility of the system, a hyperuniform system at equilibrium is effectively an incompressible system.
Theoretically, due to the fluctuation-compressibility theorem, hyperuniformity can arise naturally at thermal equilibrium in incompressible systems with long-range repulsive interactions between constituents.~\cite{Torquato2003} However, a hyperuniform point pattern can also exist within a higher-dimensional system that exhibits only short-range interactions.~\cite{Torquato2003}

In the quest for synthesizing hyperuniform materials, the impact of finite size effects is crucial to be addressed. Here we use vortex matter in type-II superconductors as a toy model system to study this issue. The nucleation of disordered hyperuniform vortex structures at the surface of samples with point disorder was reported by means of magnetic-decoration-imaging of thousands of vortices at low fields~\cite{Rumi2019,AragonSanchez2023}
and via scanning-tunnelling spectroscopy at high fields.~\cite{Llorens2019}
By using a hydrodynamic approximation to describe the interaction between vortices, some of us~\cite{Rumi2019} showed that these hyperuniform structures result from effective in-plane long-range interactions mediated by the out-of-plane elastic interaction between the tip of vortices at the surface and the body of vortices penetrating into the sample bulk.
Therefore, the hyperuniform properties of vortex matter at the surface of the samples might be affected by finite-size effects tailored for example by the thickness of the superconducting samples.~\cite{Puig2022}
Indeed,  thickness-dependent dimensional crossovers are expected in the structural properties at long wavelengths in host media with correlated planar disorder.~\cite{Puig2022,Puig2023}.

In this work we reveal the need to  avoid finite-size effects in order to  grow  novel  hyperuniform materials in host media with point disorder. We experimentally study finite-size effects in the hyperuniform vortex matter by performing magnetic decoration experiments of the vortex structure nucleated after subsequently cleaving the same Bi$_{2}$Sr$_{2}$CaCu$_{2}$O$_{8 + \delta}$ sample with weak point disorder. We perform experiments revealing vortex structures in large fields-of-view with thousands of vortices and study the thickness-dependence of the angularly-averaged
$S(q)$ in the low-$q$ limit. By discussing the variation of different metrics of the structure factor with the sample thickness, we provide experimental evidence that the hyperuniform order is worsened on decreasing the sample thickness. We discuss these results in view of the theoretical predictions provided by a hydrodynamic model of vortex matter considering effects introduced by the finite length of vortices controlled in practice by the sample thickness.

\section*{Method}

In order to study the finite-size effects in the hyperuniform properties of vortex matter, we directly image thousands of individual vortices by means of magnetic decoration experiments performed in the same Bi$_{2}$Sr$_{2}$CaCu$_{2}$O$_{8 + \delta}$ sample that was successively cleaved. Further details on the sample properties, imaging technique and methods for obtaining  the two-dimensional structure factor
$S(\mathbf{q})= S(q_{\rm x},q_{\rm y})$ of the vortex structure at the sample surface can be found in Appendix I. In order to explore how finite-size effects alter the hyperuniform properties of vortex matter, we performed decorations of freshly cleaved surfaces of the same single crystal. Studying always the same single crystal allow us to lessen the effects that can be introduced by studying samples with different levels of disorder. Our scotch-tape-based cleaving method allow us to easily prepare clean surfaces
exposed to decoration, but unfortunately the thickness of the remaining sample can not be tailored with great precision.
By successively repeating the cleaving process
we were able to vary the thickness of the original sample in 8 values ranging
from 14 to 0.5\,$\mu$m.

\section*{Results}

\begin{figure*}[ttt]
\centering
\includegraphics[width=0.93\linewidth]{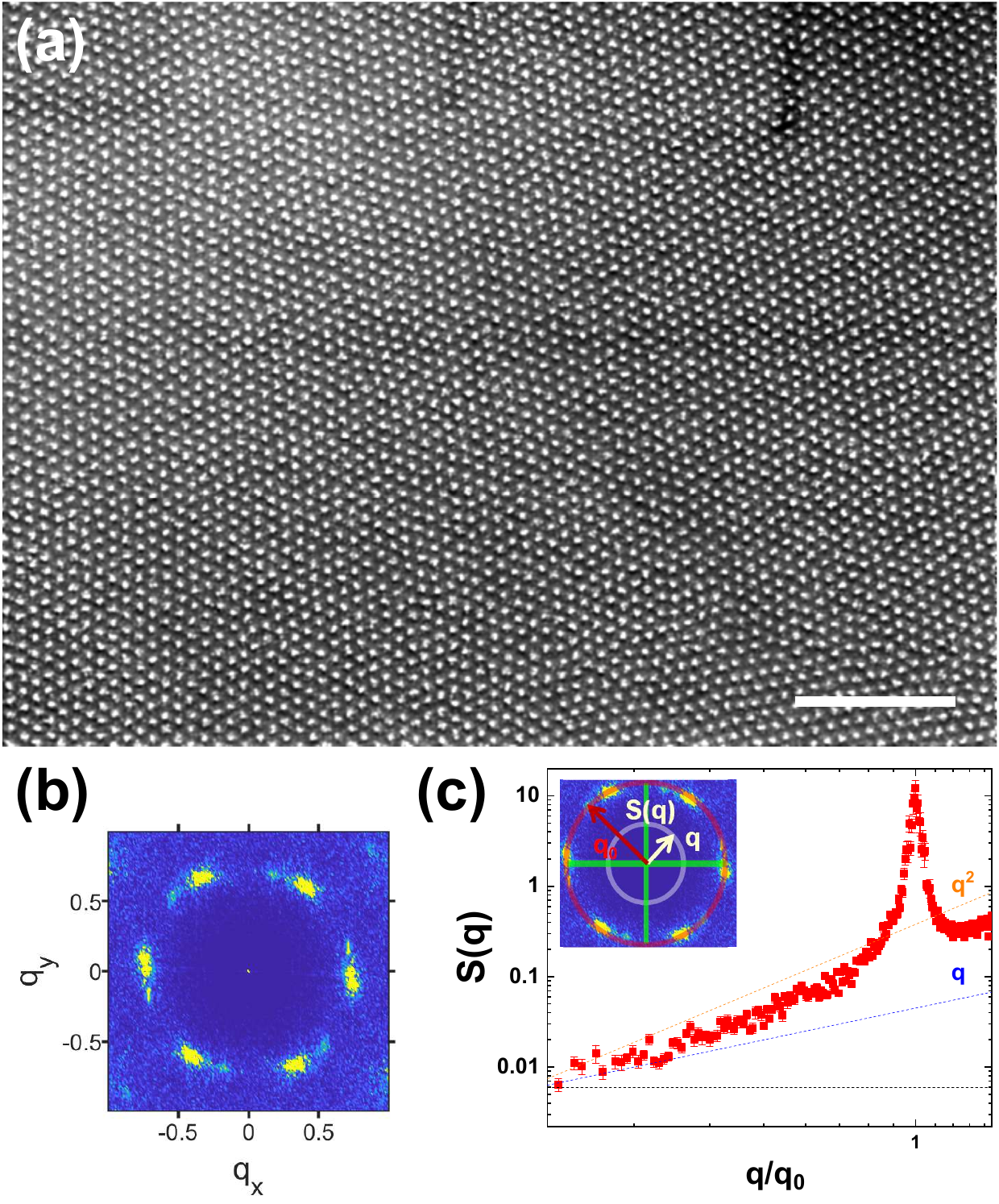}
\vspace{-0.4cm}
\caption{(a) Zoom-in of a scanning electron microscopy image of the Fe clusters (white dots) decorating vortex positions at the surface of the 14\,$\mu$m thick sample. The scale bar indicates 10\,$\mu$m. (b) Two-dimensional structure factor $S(q_{x},q_{y})$ of the vortex positions digitalized in the whole field-of-view for 14\,$\mu$m thickness. (c) Angularly-averaged structure factor $S(q)$ as a function of the wave-vector normalised by the Bragg wave-vector, $q/q_{\rm 0}$. The figure shows with dashed lines algebraic growths $q^{\alpha}$ with exponents $\alpha = 0,1,2$ (black, blue, orange) as guides to the eye. Insert: Schematics of the calculation of the angularly-averaged structure factor. The small $q$ data are affected by windowing effects and thus these points (pixels within the green cross) are not considered in the calculation of $S(q)$. }
\label{Figure1}
\end{figure*}

Figure\,\ref{Figure1} (a) shows a typical magnetic decoration image of the vortex positions at the surface of the sample for the larger studied 14\,$\mu$m thickness. Vortices are imaged as white dots and their positions are digitalized using an automatic process that searches the local maximum of intensity. In this way, we obtain the positions of the vortex tips in extended fields-of-view spanning from 4000 to 22,000 vortices depending on each experiment performed at different thicknesses. The average first-neighbors separation between vortices in all experiments is $a_{0}\sim 0.9$\,$\mu$m and for every experiment the standard deviation from this average is smaller than 10\,$\%$, suggesting vortex density fluctuations are not significant. From the digitalized vortex positions we calculate the structure factor $S(\mathbf{q} = (q_{\rm x},q_{\rm y}))$ and obtain patterns as for example the one shown in Fig.\,\ref{Figure1} (b). Irrespective of the thickness, the vortex structure presents a small number of non-sixfold coordinated vortices of roughly $2.5$\,$\%$, most of them forming edge dislocations and some of them located in grain boundaries between large crystallites with a small misalignment.  The nucleation of these small-angle grain boundaries do not significantly affect the structure factor in the low-$q$ limit.

With the aim of characterizing the vortex density fluctuations of the low-$q$ modes we compute the angularly-averaged structure factor $S(q)$, a scalar magnitude that results from averaging the structure factor within a radial section of radius $q$, see schematics in the insert to Fig.\,\ref{Figure1} (c). For our study we plot these data in a log-log scale as a function of the normalized wave-vector, $q/q_{0}$, with $q_{0}$ the Bragg wave-vector
of the hexagonal vortex structure. Figure\,\ref{Figure1} (c) shows a typical $S(q)$ for the experiment performed at 14\,$\mu$m thickness. The figure shows the experimental data in red, and dashed lines are guides to the eye representing algebraic functions $(q/q_{0})^{\alpha}$ for constant ($\alpha =0$), linear and parabolic growings. In this case the experimental structure factor shows an algebraic decay towards zero in the $q \rightarrow 0$ limit, namely $S(q) \sim q^{\alpha}$, with an exponent  between 2 and 1. Then the vortex positions at the surface of the thicker studied 14\,$\mu$m sample is a hyperuniform pattern within the type-I hyperuniformity class.~\cite{Torquato2018} The hyperuniformity class of a system is defined according to the value of the $\alpha$ exponent. Two-dimensional type-I hyperuniform patterns are the most ordered hyperuniform systems presenting $1<\alpha<2$.~\cite{Torquato2018} The hyperuniformity class can be degraded by introducing disorder~\cite{Torquato2018b,Chen2021} or a strong coupling with magneto-elastic properties of the host media.~\cite{AragonSanchez2023} We aim here to study whether the hyperuniformity class can be also depleted by  finite-size effects.

In order to systematically study this possibility, we quantify the  asymptotic density fluctuations for long wavelengths as a function of the thickness of the sample by fitting the behavior of $S(q)$ in the $q \rightarrow 0$ limit. In this limit, we have to pay attention to the windowing effect introduced by the borders and shapes of the real-space fields-of-view. In order to diminish this effect and to perform a systematic comparison between the studied thicknesses, in our analysis we cut the field-of-view in the largest rectangular window fitting the panoramic of the sample. In the case of rectangular fields-of-view, the windowing effect, an artifact associated to the Fourier transform of the edge, is manifested in $S(\mathbf{q})$ as an excess in the cross-shaped region centered around $q_{\rm x} = q_{\rm y} = 0$. This artifact is oriented along the principal directions
of the rectangle and has an oscillatory decay on increasing $q$.  In order to get rid of this spurious effect, for the smallest $q$ values we perform a partial average over the azimuth angle values but the cross-shaped region entailed by a minimal number of pixels $q_{\rm min}$ in the horizontal and vertical directions (see green cross in the insert to Fig.\,\ref{Figure1} (c)). The finite width of this exclusion cross has a safety minimal wave-vector given by the linear size of the field-of-view, $w_{\rm FOV}$, in the horizontal or vertical directions, such that $q_{\rm min} = 2\pi/w_{\rm FOV}$. Since the number of pixels to average over a fixed $q$ is a decreasing function with $q$, then the statistical fluctuations on $S(q)$ increase on going towards the $q \rightarrow 0$ limit. We can not avoid this problem by  imaging an ensemble of different vortex configurations at the same position of the sample to average over since every time a decoration experiment is performed the sample has to be cleaved as to expose a clean surface, and then we end up modifying the thickness.  Thus, even though our fields-of-view span linear sizes $w_{\rm FOV} \sim 100 a_{0}$ or more, we can safely analyze the vortex density fluctuations for $q > q_{\rm min} = 2\pi/(100 a_{0}) \sim 0.06 /a_{0}$.

\begin{figure}[ttt]
\centering
\includegraphics[width=0.72\linewidth]{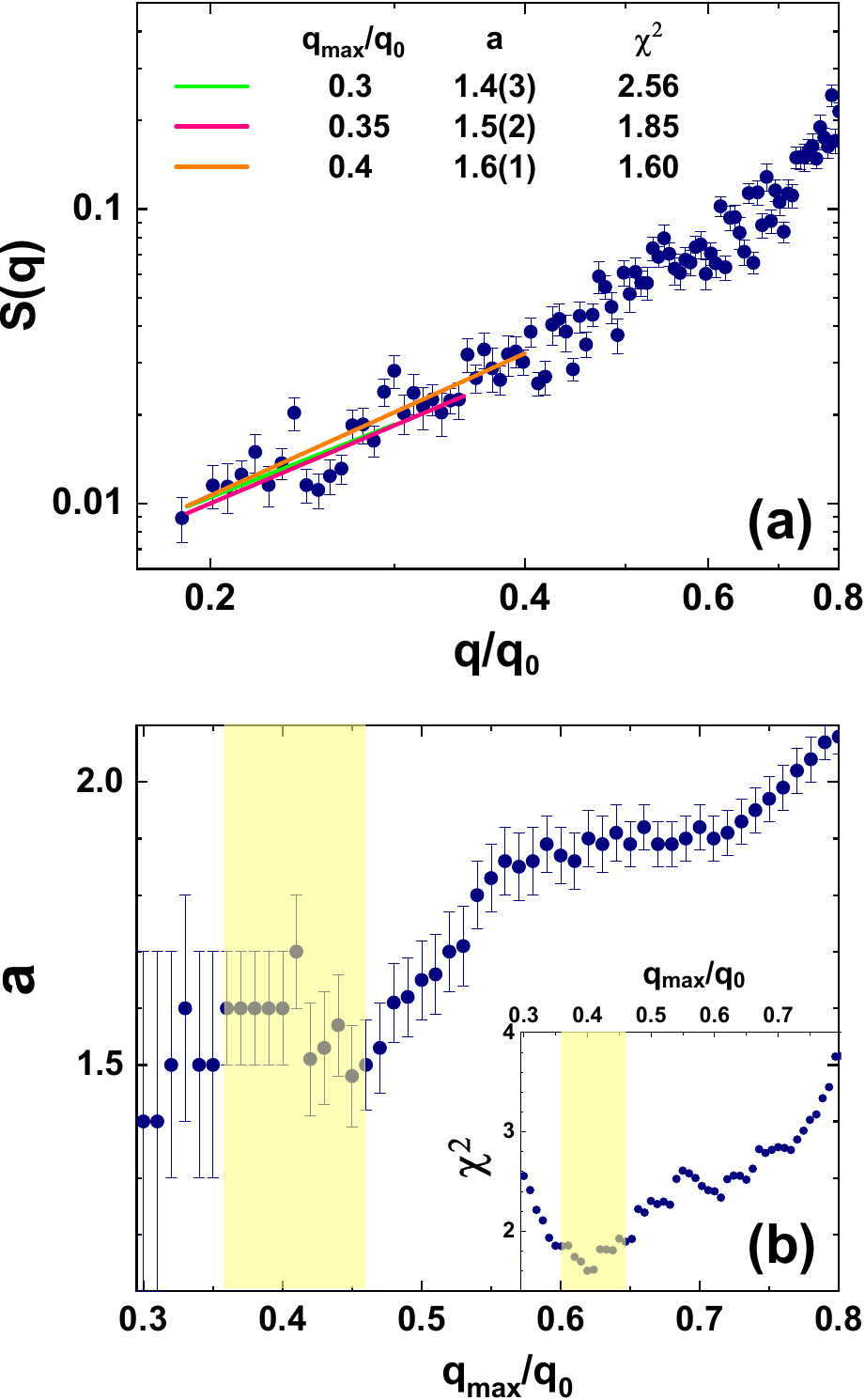}
\caption{(a) Algebraic fits (lines) of the angularly-averaged structure factor (points), $S(q) \propto q^{\rm a}$ for the vortex structure nucleated for 14\,$\mu$m thickness. The fits are performed between the minimum value of wave-vector up to variable upper bounds $q_{\rm max}/q_{0}= 0.3$, 0.35 and 0.4. The growing exponents $\rm a$ and $\chi^{2}$ confidence of the fits are indicated. (b) Main panel: Growing exponent values obtained from the fits on increasing $q_{\rm max}/q_{0}$. Insert: $\chi^{2}$ confidence of the fits. In both panels the yellowish-highlighted area indicates the range of $q_{\rm max}/q_{0}$ that we consider to obtain the effective exponent $\alpha$ as the average from the $\rm a$ values fitted for that upper bound limits range.}
\label{Figure2}
\end{figure}

\begin{figure}[ttt]
\centering
\includegraphics[width=0.68\linewidth]{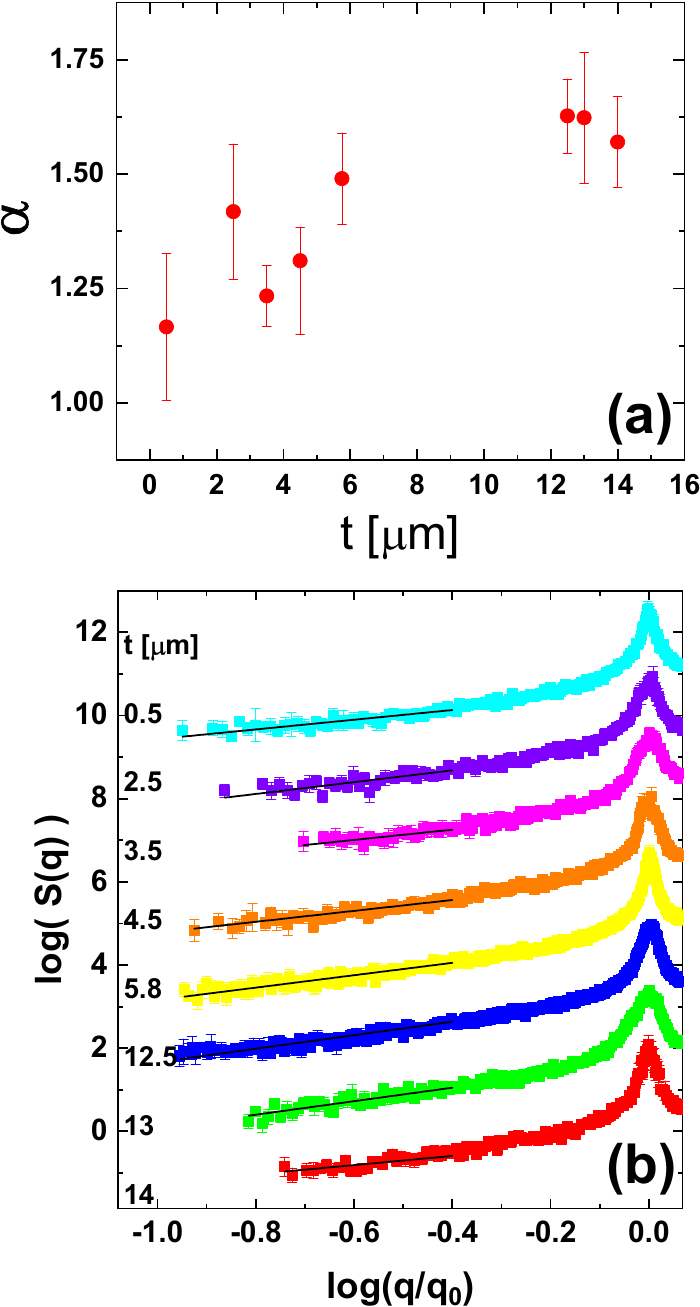}
\caption{(a) Effective $\alpha$ exponent fitted from the algebraic decrease of the angularly-averaged structure factor $S(q)$ in the $q \rightarrow 0$ limit as explained in the text. (b) Plot of the $\log(S(q))$ versus $\log(q/q_{0})$  data for all the studied thicknesses (color points) and algebraic decays $\propto (q/q_{0})^{\alpha}$ with the exponents shown in the top panel (black lines).}
\label{Figure3}
\end{figure}

\begin{figure}[ttt]
\centering
\includegraphics[width=0.68\linewidth]{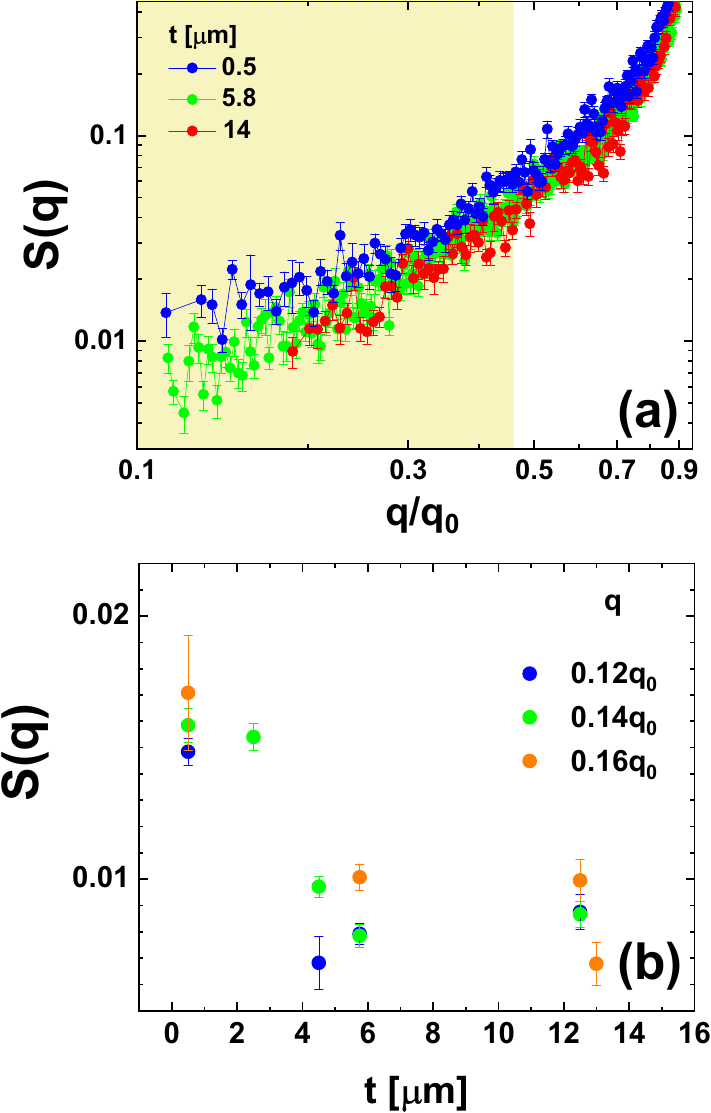}
\caption{(a) Detail of the $S(q)$ data in the low-$q$ range for the smallest, largest and intermediate thicknesses in the range studied. The yellowish-highlighted region indicates the $q$-range considered in the fits of algebraic growths of the structure factor. (b) Magnitude of the angularly-averaged structure factor for particular $q/q_{0}$ wave-vectors in the low-$q$ limit as a function of the studied thicknesses.}
\label{Figure4}
\end{figure}

We perform fits of the $S(q)$ data  following a systematic procedure that allow us to compare
data from all the studied thicknesses that start at different lower bounds for the wave-vector, $q_{\rm min}$, as well as to
improve the statistical confidence of the obtained $\alpha$ exponent. For every $S(q)$ curve obtained at every thickness $t$ we implement a series of algebraic fits $S(q) \propto q^{\rm a}$ between $q_{\rm min}/q_{0}$ and a variable upper bound for the fitting range $q_{\rm max}/q_{0}$. This upper bond varies between the value corresponding to considering 30 experimental points from $q_{\rm min}/q_{0}$ and cumulative increments of $q_{\rm max}/q_{0}$ of 0.01. For example, Fig.\,\ref{Figure2} (a) shows three fits of the 14\,$\mu$m data for $q_{\rm max}/q_{0}=0.3$, 0.35, 0.4, and their corresponding fitted  exponents, $\rm a$, and Chi-square goodness, $\chi^{2}$. Figure \,\ref{Figure2} (b) shows the variation of these two magnitudes on increasing the upper bond of the fitting window. For values of $q_{\rm max}/q_{0}<0.36$ $\rm a$ has a significant variation and large error bars as a consequence of performing fits in a window with few data. In the range $0.36 \leq q_{\rm max}/q_{0} \leq 0.46$, $\rm a$ values are more stable, their error bars are smaller, and more importantly, $\chi^{2}$ presents minimum values. On increasing $q_{\rm max}/q_{0}$ above 0.46, $\rm a$ and $\chi^{2}$ systematically increase. This is due to the upward bending of $S(q)$ induced by the development of the Bragg peak. This behavior of $\chi^2$, $\rm a$ and error bars is similar for all studied thicknesses, presenting growing $\rm a$ and $\chi^{2}$ above $q_{\rm max}/q_{0}= 0.46$.

In pursuit of obtaining a statistically
relevant exponent in the low-$q$ limit for all the studied thicknesses, we estimate the effective $\alpha$ exponent as the average of $\rm a$ values obtained in the $q_{\rm max}/q_{0}=0.36-0.46$ region, see for example the yellowish-highlighted areas of Fig.\,\ref{Figure2} (b). The absolute value of $\alpha$ obtained in this way may not accurately represent the asymptotic value of the exponent that describes the algebraic slow-down of density fluctuations at large wavelengths, but allows us to perform a systematic comparison between data obtained at different thicknesses for experiments with different values of $q_{\rm min}/q_{0}$.

The effective exponents obtained following this procedure are shown in Fig.\,\ref{Figure3} for all the studied thicknesses. Panel (a) presents the main result of this work, namely that $\alpha$ decreases when the sample gets thinner. Panel (b) shows that the algebraic growths $\propto (q/q_{0})^{\alpha}$ (black lines) follow the experimental data (colour points) in the low-$q$ region. These results imply that finite-size effects alter the hyperuniform properties of vortex matter for this range of thicknesses studied: Even though hyperuniformity remains type-I in this $t$-range down to 0.5\,$\mu$m, there is a tendency towards an enhancement of density fluctuations at long distances when decreasing the sample thickness. How $\alpha$ depends on $t$ deserves further investigation considering a realistic theoretical model of elastic interactions in this system, but this is beyond the scope of this work. Nevertheless, experimentally we observe that the hyperuniform hidden order is systematically depleted when $t$ decreases.

The latter finding is also supported by the data shown in Fig.\,\ref{Figure4} (a) presenting the structure factor  in the $q \rightarrow 0$ limit for the smallest, largest and an intermediate $t$  from the range studied. In the $q \rightarrow 0$ limit, the $S(q)$ values roughly increase with decreasing thickness. This is also evident from the data of panel (b) showing the variation with thickness of $S(q)$ computed for particular low-$q/q_{0}$ values. For instance, for the small $q/q_{0} =0.12$ value the magnitude of $S(q)$ increases roughly a factor 2 when the thickness varies from the largest to the smallest studied.
The data shown in panel (a) also suggests that the $S(q)$ curves for small $t$ depart upwards from curves obtained at larger $t$ at a certain critical value of $q/q_{0}$ that seems to increase on decreasing thickness, see forking of the three curves for small $q$. For wave-vectors smaller than these critical values the $S(q)$ curve seems to bend from the algebraic behavior observed at larger $q$, following a growth with a lower exponent. For the smallest thicknesses the data even show a tendency towards saturation at the lowest accessed $q$, see for example the data at the smallest thickness of 0.5\,$\mu$m for $q/q_{0}< 0.15$. It is important to point out that the range of $q_{max}/q_{0}$ up to which the fits are performed is well above these critical $q$ values, see yellowish-highlighted region in Fig.\,\ref{Figure4} (a) indicating the fitting region. Nevertheless, the decrease of $\alpha$ with $t$ presented in Fig.\,\ref{Figure3} can be also underestimated by the fitting procedure we are using, forced by the fact that the low-$q$ data are noisy since when performing the angular average the statistical frequency decreases. Irrespective of this detail, the observed upwards bending of the $S(q)$ curves at critical $q$ values is another manifestation that on decreasing $t$ finite-size effects play a significant role on depleting the hyperuniform properties of an elastic system.

\section*{Discussion}

As shown in the previous section, with decreasing thickness the structure factor decreases systematically slower when $q \rightarrow 0$, but for the studied range $\alpha > 1$. These effective exponents found experimentally contrast with the exponent equal to zero predicted theoretically for the solid Bragg glass phase within a hydrodynamic approximation.~\cite{Rumi2019}
Since here we discuss the experimental observations by considering the predictions of this hydrodynamic model that we presented in a previous work,~\cite{Rumi2019} Appendix II presents a detailed description of this model.
 The Bragg glass  presenting quasi-long-range positional order is the stable vortex solid phase expected~\cite{Giamarchi1995,Klein2001} for
Bi$_{2}$Sr$_{2}$CaCu$_{2}$O$_{8 + \delta}$ samples with weak point disorder as the ones studied here. As some of us pointed out in previous works,~\cite{Rumi2019,Puig2022} this discrepancy is due to the fact that the hydrodynamic approximation is an equilibrium prediction but, during the experimental cooling protocol non-equilibrium effects  play an important role. The vortex structure observed at low temperatures after a field-cooling protocol is an out-of-equilibrium structure since the relaxation rate of the structure strongly depends on the wavelength or wave-vector of the density fluctuation modes. Low-$q$ density fluctuations have a slower relaxation rate than large-$q$ density fluctuations and then the vortex structure gets frozen at a freezing temperature that depends on the observation lengthscale,  namely $T_{\rm freez} = T_{\rm freez}(q)$.   The local ordering of the structure at lengthscales of $a_{0}$ indicates that $T_{\rm freez} ( q \sim  1/a_{0}) = T_{\rm irr} \sim T_{\rm m}$, the melting temperature of the vortex structure.~\cite{CejasBolecek2016}  Density fluctuations that determine the $S(q)$ behavior for $q \rightarrow 0$ fall out of equilibrium at higher temperatures, namely
$T_{\rm freez} ( q \rightarrow 0)$ is probably larger than the melting temperature.
 In addition, this out-of-equilibrium effect is expected to be enhanced by disorder that slows down significantly the thermally-activated dynamics.
Then, the $q \rightarrow 0$ modes retain memory of the liquid phase that is a hyperuniform state.
This might explain our experimental observation of an $\alpha > 0$ instead of the theoretically expected value of an algebraic growing exponent equal to zero  for the solid Bragg glass phase. Since for a liquid phase $\alpha=1$,~\cite{Rumi2019} the measured $\alpha>1$ value might be associated to the dispersive nature of the elastic constants  of a highly-anisotropic vortex structure such as that of Bi$_{2}$Sr$_{2}$CaCu$_{2}$O$_{8 + \delta}$. The dispersivity of elastic constants tends to increase the non-dispersive, $S(q)\sim q$, growth of density fluctuations with a functionality $S(q)\sim q^2$ on further increasing the wave-vector $q$.

\begin{figure}[ttt]
\centering
\includegraphics[width=\linewidth]{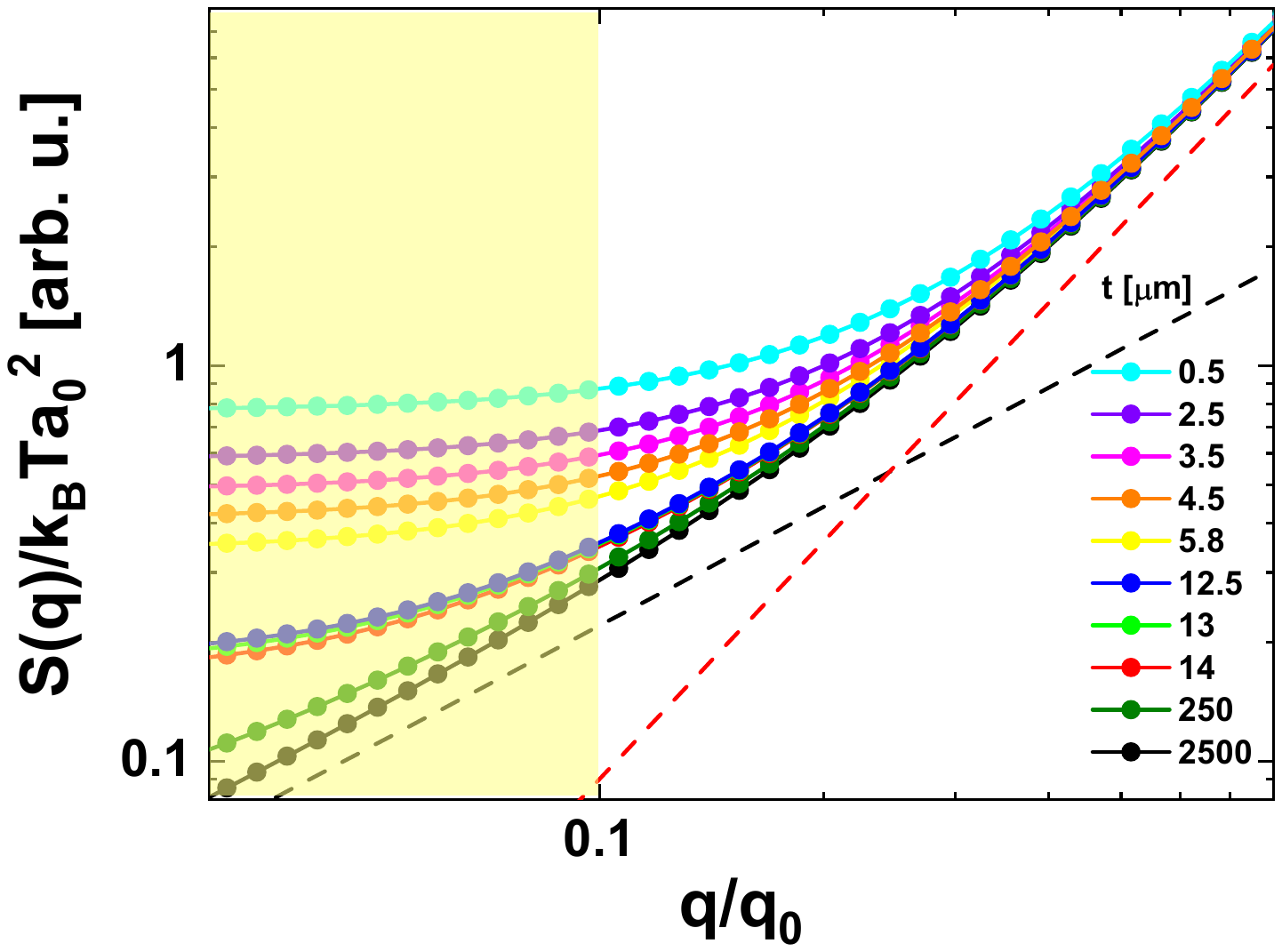}
\caption{Theoretical angularly-averaged structure factor in the low-$q$ limit at the surface of the sample for the
three-dimensional vortex line liquid without disorder within the
hydrodynamic approximation. Data are calculated for the same vortex density as in the experiments reported here ($a_{0} = 0.9$\,$\mu$m) and for superconducting parameters corresponding to the highly anisotropic Bi$_{2}$Sr$_{2}$CaCu$_{2}$O$_{8 + \delta}$ system, see text for further details. Different colors represent data for the different thicknesses $t$ studied experimentally (0.5 to 14\,$\mu$m) and larger (250 to 2500\,$\mu$m). Dashed lines are guides to the eye corresponding to linear (black) and quadratic (red) algebraic growths. The $q$-range with white background ($0.1-0.8$) corresponds to data accessed experimentally whereas the range with yellow background corresponds to the theoretical prediction beyond typical fields-of-view.}
\label{Figure5}
\end{figure}

 As some of us already discussed,~\cite{Rumi2019} finite-size effects can induce a crossover from hyperuniformity to non-hyperuniformity when decreasing $q$. In order to quantify this finite-size effect and compare with our experimental observations, the strong dispersivity of elastic constants in the studied vortex structure has to be considered.  The dispersion of the elastic constants $c_{\rm 11}$ and $c_{\rm 44}$ in the $z$-direction becomes important if
$q \lambda_{\rm c} \gtrsim 1$, with
$\lambda_{\rm c}$ the penetration depth for supercurrents running in $c$-axis direction (field applied in the $ab$ plane). Since in the studied samples the superconducting anisotropy $\Gamma = \lambda_{\rm c}/\lambda_{\rm ab} = 170$, then $\lambda_{\rm c} = \Gamma \lambda_{\rm ab} = 170$\,$\mu$m$\sim 180 a_{0}$ and the condition
$q \lambda_{\rm c} \sim 1$  implies that dispersion effects are dominant for $q/q_{0} \gtrsim 10^{-3}$. Therefore for the lowest $q$ accessed in our experiments dispersion effects start to become relevant. This can be analytically calculated  starting with the expression of the structure factor expected for the liquid or solid vortex phases with no disorder at thermal equilibrium at temperature $T$  in the hydrodynamic approximation,

\begin{align}
S^{3d}_{\rm sol} (q, q_{\rm z} ) = S^{3d}_{\rm liq} (q, q_{\rm z} )= \frac{n_{0}k_{\rm B}Tq^{2}}{q^{2}c_{\rm 11}(q, q_{\rm z} ) + q^{2}_{z} c_{\rm 44}(q,q_{\rm z} )},
\label{eq:s3dq}
\end{align}

\noindent and then replacing in the latter equation the $q$ and $q_{\rm z}$-dependent dispersive compressive and tilting elastic constants that at in the lowest order approximation and for low wave-vectors have the expressions~\cite{Blatter1994}

\begin{align}
c_{\rm 11} (q,q_{\rm z}) = \frac{B^{2}}{4\pi} \frac{1 + \lambda_{\rm c}^{2}(q^{2} + q_{\rm z}^{2})}{(1 + \lambda_{\rm ab}^{2}(q^{2} + q_{\rm z}^{2}))(1 + \lambda_{\rm c}^{2}q^{2} + \lambda_{\rm ab}^{2}q_{\rm z}^{2})}
\label{eq:c11}
\end{align}

\begin{align}
c_{\rm 44} (q,q_{\rm z}) = \frac{B^{2}}{4\pi} \frac{1}{1 + \lambda_{\rm c}^{2}q^{2} + \lambda_{\rm ab}^{2}q_{\rm z}^{2}} + c'_{\rm 44} (q_{\rm z})
\label{eq:c44}
\end{align}

\noindent with $c^{'}_{\rm 44} (q_{\rm z})$ the contribution from an isolated vortex that in the short wave-vector limit we are studying can be considered as non-dispersive.~\cite{Blatter1994} In order to obtain the two-dimensional structure factor at the surface of the system, $S(\mathbf q)$, the $S^{3d} (\mathbf q, q_{\rm z} )$ has to be summed up the discrete values of $q_{\rm z}$ given by the thickness of the sample such that the modes of the perturbation in the $z$ direction can only have values $q_{\rm z}=2 \pi n/ t$,  namely

\begin{align}
    S(\mathbf q)
    \propto \sum_{n=0}^{n_{\rm max}}
    S^{3d} (\mathbf{q}, q_{\rm z}= 2\pi n/t ).
    \label{eq:fromS3dtoS2d}
\end{align}
\noindent where $n_{\rm max} \approx t/s$ with $s\ll t$ represents a short-range cut-off in the $z$-direction that can be set to the  superconductor layer spacing. Since the sum converges rapidly, the exact value of $n_{\rm max}$ is irrelevant, providing it is large enough. Then, from this magnitude we compute the angularly-averaged value $S(q)$ at equilibrium within the hydrodynamic approximation
that can be compared with the experimental data. We performed these calculations taking into account the experimental parameters $\lambda_{\rm ab}(T_{\rm freez}) = 0.4$\,$\mu$m,   $a_{0}/\lambda_{\rm ab}(T_{\rm freez})=2$, and the eight thicknesses studied experimentally. The result of this calculation is presented in Fig.\,\ref{Figure5} where we have also included results for thicker samples and for a larger range of $q\rightarrow 0$ with respect to the one explored experimentally in order to highlight some asymptotic trends.
From Eq.\ref{eq:s3dq} we expect the theoretical $S(q)$ to saturate at a positive value for $q<q_{\rm FS}$ where

\begin{align}
q_{\rm FS} \approx \frac{2\pi}{t}\sqrt{\frac{c_{44}}{c_{11}}}
\end{align}

\noindent with $c_{44}\approx c_{44}(q_{\rm FS},2\pi/t)$
and $c_{11}\approx c_{11}(q_{\rm FS},2\pi/t)$ from Eqs. \ref{eq:c11} and \ref{eq:c44}. Hence, the saturation is a finite-thickness effect that enhances on decreasing the thickness of the sample, as can be appreciated in Fig.\,\ref{Figure5}. In addition, for a large enough value of $t$ and for $q > q_{\rm FS}$, a second crossover takes place. This crossover is dictated by the dispersivity of the elastic constants $c_{11}$ and $c_{44}$, between a non-dispersive $S(q)\sim q$ to a dispersive $S(q)\sim q^2$ regime on increasing $q$. Since $q_{FS}\to 0$ as $t\to \infty$, this implies that infinitely thick samples are expected to be class-II hyperuniform with $\alpha > 1$. However, for small $t$ the $q$-range of the $S(q)\sim q$ regime can be significantly reduced and can even become undetectable experimentally. In Fig.\,\ref{Figure5} we can appreciate these effects by comparing $t=0.5\;\mu$m where a clear $S(q)\sim q$ regime is visible before saturation by further decreasing $q$, and $t=2500\;\mu$m, where a $S(q)\sim q^2$ crosses over directly to saturation by decreasing $q$.

As previously mentioned, the theoretically predicted behavior of $S(q)$ showing a tendency towards saturation at small $q$ on decreasing thickness, with the saturation value increasing with decreasing $t$, is also observed experimentally. Moreover, the relative change of $S(q)$ for $q/q_{0} =0.12$ for $t=0.5$\,$\mu$m than for 14\,$\mu$m is theoretically of order $\approx 2$, in striking agreement with the experimental.  Therefore, the phenomenology observed in the low-$q$ limit of the experimental curves can be roughly explained by the finite thickness of the sample.

The experimentally observed variation with thickness of the effective algebraic exponent $\alpha$ describing the suppression of density fluctuations at large scales can also be understood considering these hydrodynamic calculations.  When $t$ is large the saturation
associated to finite-size effects is well below the $q$-range considered to fit $\alpha$ (see yellowish-highlighted area in Fig.\,\ref{Figure4} (a)). Nevertheless, if the thickness is sufficiently decreased the finite-size effects produce the upward bend of $S(q)$ at small $q$ and thus effectively diminishes $\alpha$ for the whole fitting range. Therefore, this crossover between finite-size effects when $q \rightarrow 0$ and algebraic growing of $S(q)$ at small-intermediate $q$  explains why the $\alpha$ estimated in the $q/q_{0}$ range up to 0.46 lessens when decreasing thickness. Then the experimental $\alpha$ results from fitting the $S(q)$ data in a low-$q$ range where takes place a trade-off between a tendency to saturation due to finite-size effects and an algebraic growth with exponent larger than one due to the dispersivity of elastic constants in a highly anisotropic vortex system.

\section*{Conclusion}

We present experimental evidence on the degradation of hyperuniformity in vortex matter induced by a finite-size effect. This degradation is manifested in two phenomenological observations that are inter-twinned: A diminishing of the effective algebraic exponent describing the decrease of density fluctuations  at large wavelengths and a crossover towards saturation of density fluctuations for wavelengths larger than a critical value that shortens with decreasing the thickness of the system. The last phenomenology is a description of the evolution of density fluctuations with wave-length
considering the observed behavior of $S(q)$ in the reciprocal wave-vector space obtained from real-space images of vortex matter in extended fields-of-view for the same Bi$_{2}$Sr$_{2}$CaCu$_{2}$O$_{8 + \delta}$ sample with different thickness. Then, by considering a theoretical hydrodynamic model and the dispersivity of the elastic constants for the anisotropic elastic system studied, as well as considering that the small-$q$ modes of the structure get frozen at temperatures of the order of the melting temperature,  we show how varying the thickness of the sample can induce the experimentally-observed depletion of hyperuniformity on decreasing $t$.  In conclusion, our work makes the important contribution that finite-size effects can eventually degrade the hyperuniform properties of the system of interacting objects. These effects have to be considered seriously when engineering hyperuniform materials for relevant technological applications.

\section*{Acknowledgements}We thank W. Turmbr{\"a}u for insightful discussions. Work supported by the National Council of Scientific and Technical Research of Argentina (CONICET)
through grant PIP 2021-1848, by the  Argentinean Agency for the Promotion of Science and Technology (ANPCyT) through grants PICT 2018-1533 and PICT 2019-1991, and by the Universidad
Nacional de Cuyo research Grants 06/C008-T1
and 06/C014-T1. Y. F. thanks funding from the Alexander von Humboldt Foundation through the Georg Forster Research Award and from the Technische Universit\"{a}t Dresden through the Dresden Senior Fellowship Program.

\section*{Appendix I: Sample, experimental, and analysis details}

The sample studied here is a nearly optimally-doped single crystal with $T_{\rm c} \sim 90$\,K, grown by means of the flux method.~\cite{Correa2001} The sample was specially selected since it does not present any visible planar defect as revealed by magnetic decoration.  Thus the dominant disorder in this particular sample is point like, natural atomic-scale defects distributed at random that arise when growing the crystals.

The magnetic decoration is an imaging technique that reveals the positions of individual vortices at the sample surface. This is performed by  evaporating ferromagnetic particles that are attracted towards the magnetic halo of vortices where the local magnetic field presents a maximum that decays with distance in lengthscales of the superconducting penetration depth $\lambda_{\rm ab}$.~\cite{Fasano2008} Thus the ferromagnetic particles evaporated in the sample surface decorate the positions of the vortex tips. This technique can be applied to directly image vortex positions in the whole millimeter-size sample surface, allowing for the identification of the vortex positions at the surface in fields-of-view spanning thousands of vortices.  By means of scanning-electron-microscopy, panoramic views of the vortex structure can be obtained after a digitalization procedure of the images of the sample surface with ferromagnetic particles decorating the vortex positions.~\cite{Fasano2003} In order to study vortex density fluctuations at large lengthscales, magnetic decoration imaging is best suited than other techniques such as scanning-tunnelling spectroscopy,~\cite{Llorens2019,Petrovic2009} magnetic force or scanning-squid microscopy~\cite{Llorens2020} that typically image only up to hundreds of vortices.

Here we present data for a magnetic induction of 30\,G providing $a_{0} \sim 0.9$\,$\mu$m. The experiments were performed at 4.2\,K after following a field-cooling protocol from the normal state as described elsewhere.~\cite{Fasano1999} During this experimental protocol the
vortex structure gets frozen at lengthscales of $a_{0}$
at a temperature $T_{\rm freez}$ that is very close to the irreversibility temperature where the pinning associated with the sample disorder sets in.~\cite{CejasBolecek2016,Fasano2005}
When further cooling down $T_{\rm freez}$, vortices profit from disorder making local excursions on lengthscales of roughly coherence length, two orders of magnitude smaller than the spatial resolution of the decoration technique $\sim\lambda_{\rm ab} (0) \sim 0.2$\,$\mu$m$ \sim a_{0}/4$.
The irreversibility temperature for near optimally-doped samples as the one studied here ($T_{\rm c} \sim 90$\,K), and for an induction of  $\sim 30$\,G, is around 85\,K.~\cite{Correa2001,Dolz2014,Dolz2015}
Therefore the panoramics obtained by magnetic decoration at 4.2\,K reveals the local density fluctuations at lengthscales of
$ a_{0}$ for the vortex structure frozen at $T_{\rm freez}\sim 85$\,K.

Once vortex positions are digitalized from the panoramic images, we obtain the two-dimensional structure factor
$S(\mathbf{q})= S(q_{\rm x},q_{\rm y})$ of the vortex structure at the sample surface by Fourier-transforming the local vortex density fluctuations.
This experimental approach is complementary to the small-angle-neutron-scattering method that allows to directly measure the diffracted neutron intensity proportional to the structure factor of the vortex lattice along the bulk of the sample.~\cite{AragonSanchez2012} This latter technique lacks good resolution at small wave-vectors.~\cite{AragonSanchez2012} On the contrary, since magnetic decoration allows to image extended fields-of-view, the low-$q$ limit can be accessed with reasonable resolution. Magnetic decoration discerns individual vortex positions at low vortex densities  given by the average vortex spacing $a_{0} \propto 1/\sqrt{B}$.

\section*{Appendix II: Hydrodynamic Model}

Since we are interested in modeling the large-wavelength density fluctuations of the vortex structure, a coarse-grained description is quite convenient. A good starting point for such hydrodynamic description is the Landau free energy functional, $F$, describing $N$ interacting and elastic lines directed along the $z$ direction~\cite{Nelson1990,Marchetti1991} as

\begin{eqnarray}
F&= \frac{1}{2 n_{0}^{2}} \int d^{2} \mathbf{r} d z \int d^{2}
\mathbf{r}^{\prime} \int d z^{\prime}
[c_{44}\left(\mathbf{r}-\mathbf{r}^{\prime}, z-z^{\prime}\right)
\mathbf{t}\left(\mathbf{r}, z\right) \cdot \mathbf{t}\left(\mathbf{r}^{\prime}, z^{\prime}\right) \nonumber \\
&+c_{11}\left(\mathbf{r}-\mathbf{r}^{\prime}, z-z^{\prime}\right)
\delta n\left(\mathbf{r}, z\right) \delta
n\left(\mathbf{r}^{\prime}, z^{\prime}\right)]
+\int d^{2} \mathbf{r} \int d z V_{D}(\mathbf{r},z) \delta
n(\mathbf{r}, z) \label{eq:freeenergy}
\end{eqnarray}

\noindent where $\mathbf{r}=(x,y)$. The non-local tilt and compression modulii,
$c_{44}\left(\mathbf{r}-\mathbf{r}^{\prime}, z-z^{\prime}\right)$
and $c_{11}\left(\mathbf{r}-\mathbf{r}^{\prime},
z-z^{\prime}\right)$, have Fourier transforms
$c_{11}(\qperp,q_z)$ and $c_{44}(\qperp,q_z)$.
The two-dimensional vortex density fluctuations around its mean value $n_0$ at a layer located at
$z$ is

\begin{equation}
\delta n\left(\mathbf{r}, z\right)=\sum_{j=1}^{N}
\delta\left[\mathbf{r}-\mathbf{r}_{j}(z)\right]-n_0.
\end{equation}

\noindent The two-dimensional tangent field density for a collection
of $N \gg 1$ vortex-lines positioned at $\mathbf{r}_{j}(z)$ at a
given constant-$z$ cross section is then

\begin{equation}
\mathbf{t}\left(\mathbf{r}, z\right)=\sum_{j=1}^{N} \frac{d
\mathbf{r}_{j}}{d z} \delta\left[\mathbf{r}-\mathbf{r}_{j}(z)\right].
\end{equation}

\noindent The tangent and density fields are not independent but related by the continuity equation~\cite{Nelson1990}

\begin{equation}
\partial_{z} \delta n+\nabla_{\perp} \cdot \mathbf{t}=0 .
\label{eq:conservation}
\end{equation}

\noindent If we assume a slab geometry with a
thickness $L$ in the $z$-direction and an area $A$ in the $x-y$
plane, we get $n_0 = N/A \equiv B/\Phi_0$. The last term in $F$ describes the
coupling of the vortex density with the pinning potential
$V_{D}(\mathbf{r},z)$, which can have different correlations.

The main quantity we are interested in is the density fluctuation in Fourier space, $\delta n({\bf q},q_z)$,
from which the full three-dimensional structure factor
\begin{equation}
n_0 S^{3d}(\bqperp,q_z) = \overline{\langle |\delta
n(\bqperp,q_z)|^2 \rangle} \label{eq:defSofq3d}
\end{equation}
can be obtained, with the brackets indicating average over thermal fluctuations with the partition function $e^{-F/k_B T}$ and the
overline average stands over disorder realization. Solving the complete problem to compute $S^{3d}(\bqperp,q_z) $ is a formidable task due to the presence of disorder. Nevertheless, solutions in the absence of disorder are still useful to interpret field-cooling experiments where memory of the high-temperature liquid phase is expected to be retained at low temperatures for large-wavelengths, coexisting with local thermal equilibration.

In the absence of disorder, $F$ becomes quadratic in $\delta n({\bf r},z)$ and ${\bf t}({\bf r},z)$, and Fourier modes become decoupled. Using the continuity equation, expressed as $q_z \delta n(\mathbf{q},q_z)+\mathbf{q} \cdot \mathbf{t}(\mathbf{q},q_z)=0$ in Fourier space, we obtain a closed description for $\delta n(\mathbf{q},q_z)$. Then, applying the convolution theorem in (\ref{eq:freeenergy}) and the energy equipartition theorem, we straightforwardly derive
\begin{equation}
{S}^{3d}_{\text{liq}}(\bqperp,q_z) \approx
S^{3d}_{\text{sol}}(\bqperp,q_z) = \frac{n_0 k_B T q^2}{q^2
c_{11}(\qperp, q_z) +q_z^2 c_{44}(\qperp, q_z)}.
\label{eq:Sofq3dliquid}
\end{equation}
Assuming translation invariance along $z$ and using Eq.(\ref{eq:fromS3dtoS2d}) we can access to the two dimensional structure factor $S({\bf q})$ describing density fluctuations for a constant $z$ plane. It is important to point out that the results depend on the precise functional form of the dispersive elastic constants $c_{11}(q,q_z)$ and $c_{44}(q,q_z)$.
Only small corrections to $S({\bf q})$ are expected if more realistic open boundaries conditions are considered.~\cite{Marchetti1991}.


\section*{References}

\begin{thebibliography}{<99>}

\bibitem{Torquato2003} Torquato S and  Stillinger F H 2003 {\it Phys. Rev. E} \textbf{68} 041113.

\bibitem{Man2013} Man W, Florescu M, Williamson E P,  He Y, Reza Hashemizad S, Leung B Y C, Liner D R,  Torquato S, Chaikin P M and Steinhardt PJ 2013 {\it Proc. Natl. Acad. Sci.} \textbf{110} 15886.

\bibitem{Torquato2018} Torquato S 2018 {\it Phys. Rep.} \textbf{745}  1.

\bibitem{Chen2018} Chen D and Torquato S 2018 {\it Acta Materialia} \textbf{142} 152.



\bibitem{Zheng2020} Zheng Y, Liu L, Nan H, Shen Z -X, Zhang G, Chen D, He L, Xu W, Chen M, Jiao Y and  Zhuang H 2020, {\it Sci. Adv.} \textbf{6} eaba0826.

\bibitem{Florescu2009} Florescu M, Torquato S and Steinhardt P J 2009 {\it Proceed. Nat. Acad. Sci.} \textbf{106} 20658.

\bibitem{Froufe2016} Froufe-P{\'e}rez L S, Engel M, Damasceno P S, Muller N, Haberko J, Glotzer S C and Scheffold F 2016 {\it Phys. Rev. Lett.} \textbf{117} 053902.

\bibitem{Rumi2019} Rumi G, Aragón Sánchez J, Elías F,  Cortés Maldonado R, Puig J, Cejas Bolecek N R, Nieva G, Konczykowski M, Fasano Y and Kolton A B 2019 {\it Phys. Rev. Res.} \textbf{1} 033057.

\bibitem{AragonSanchez2023} Aragón Sánchez J, Cortés Maldonado R, Amigó M L, Nieva G, Kolton A B and Fasano Y 2023, {\it Phys. Rev. B} \textbf{107} 094508.

\bibitem{Llorens2019} Llorens J B, Guillamón I, Serrano I G, Córdoba R, Sesé J, De Teresa J M, Ibarra M R, Vieira S, Ortuño M and Suderow H 2020 {\it Phys. Rev. Res.} \textbf{2} 033133.

\bibitem{Puig2022} Puig J, Elías F, Aragón Sánchez J, Cortés Maldonado R, Rumi G, Nieva G, Pedrazzini P, Kolton A B and Fasano Y 2022 {\it Commun. Mater.} \textbf{3} 32.

\bibitem{Puig2023} Puig J,  Aragón Sánchez J, Nieva G,  Kolton A B and Fasano Y 2023 {\it Journal Phys. Soc. Japan: Conf. Proceed.} \textbf{38} 011051.

\bibitem{Torquato2018b} Kim J and Torquato S 2018 {\it Phys. Rev. B} \textbf{97} 054105.

\bibitem{Chen2021} Chen D, Zheng Y, Lee C-H, Kang S, Zhu W, Zhuang H, Huang P Y and Jiao Y 2021 {\it Phys. Rev. B} \textbf{103} 224102.


\bibitem{Giamarchi1995} Giamarchi T and Le Doussal P 1995 {\it Phys. Rev. B} \textbf{52} 1242.

\bibitem{Klein2001} Klein T, Joumard I, Blanchard S, Marcus J, Cubitt R, Giamarchi T and Le Doussal P 2001 {\it Nature} \textbf{413} 404.


\bibitem{CejasBolecek2016} Cejas Bolecek N R, Kolton A B, Konczykowski M, Pastoriza H, Dominguez D and  Fasano Y 2016 {\it Phys. Rev. B} \textbf{93} 054505.


\bibitem{Blatter1994} Blatter G, Feigel’man M V, Geshkenbein V B, Larkin A I and Vinokur V M 1994 {\it Rev. Mod. Phys.} \textbf{66} 1125.

\bibitem{Correa2001} Correa V F, Kaul E E and Nieva G 2001 {\it Phys. Rev. B} \textbf{63} 172505.

\bibitem{Fasano2008} Fasano Y and Menghini M, {\it Supercond. Science and Tech.} 2008 \textbf{21} 023001.


\bibitem{Fasano2003} Fasano Y, De Seta M, Menghini M, Pastoriza H and De la Cruz F 2003 {\it Solid State comm.} \textbf{128} 51.


\bibitem{Petrovic2009}  Petrović AP, Fasano Y, Lortz R, Senatore C, Demuer A, Antunes AB, Paré A, Salloum D, Gougeon P, Potel M,  Fischer Ø 2009 {\it Phys. Rev. Lett.} \textbf{103} 257001.

\bibitem{Llorens2020} Llorens J B, Embon L, Correa A, González J D, Herrera E, Guillamón I, Luccas R F, Azpeitia J, Mompeán F J, García-Hernández M, Munuera C, Aragón Sánchez J, Fasano Y, Milošević M V, Suderow H and Anahory Y 2020 {\it Phys. Rev. Res.} \textbf{2} 013329.



\bibitem{Fasano1999} Fasano Y, Herbsommer J and De la Cruz F 1999 {\it Phys. Stat. Sol.} \textbf{215} 563.

\bibitem{Fasano2005} Fasano Y, De Seta M, Menghini M, Pastoriza H and De La Cruz F 2005 {\it Proceed. Nat. Acad. Sci.} \textbf{102} 3898.



 \bibitem{Dolz2014} Dolz M I, Fasano Y, Pastoriza H, Mosser V, Li M and Konczykowski M 2014 {\it Phys. Rev. B} \textbf{90} 144507.

\bibitem{Dolz2015} Dolz M I, Fasano Y, Cejas Bolecek N R, Pastoriza H, Mosser V, Li M and Konczykowski M 2015 {\it Phys. Rev. Lett.} \textbf{115} 137003.

\bibitem{AragonSanchez2012} Aragón Sánchez J, Cortés Maldonado R, Cejas Bolecek N R, Rumi G, Pedrazzini P, Dolz M I, Nieva G, van der Beek C J, Konczykowski M, Dewhurst C D, Cubitt R, Kolton A B, Pautrat A and Fasano Y 2012 {\it Comm. Phys.} \textbf{2}  143.

\bibitem{Nelson1990}  Nelson D R and Le Doussal P 1990 {\it Phys. Rev. B} \textbf{42} 10113.

\bibitem{Marchetti1991} Marchetti M C and  Nelson D R 1991 {\it Phys. C} \textbf{174} 40.

\end{thebibliography}
\end{document}